\documentclass[twocolumn,aps,prc,showpacs,superscriptaddress,preprintnumbers,floatfix,nofootinbib]{revtex4}
\usepackage{epsfig,graphics}
\usepackage{graphicx}
\usepackage{dcolumn}
\usepackage{bm}
\usepackage{amsmath}
\usepackage[usenames]{color}
\usepackage{ulem} 
\usepackage[colorlinks,linkcolor=blue,urlcolor=blue,citecolor=blue]{hyperref}

\begin{document}

\title{Isoscalar giant monopole resonance in Sn isotopes using a quantum molecular dynamics model}

\author{ C. Tao}
\thanks{Present address: Shandong Tumor Hospital, Jinan 250117, China}
\affiliation{Shanghai Institute of Applied Physics, Chinese
Academy of Sciences, Shanghai 201800, China}
\author{ Y. G. Ma}
\thanks{Corresponding author: ygma@sinap.ac.cn}
\affiliation{Shanghai Institute of Applied Physics, Chinese Academy of Sciences, Shanghai 201800, China}
\affiliation{Kavli Institute for Theoretical Physics, Chinese Academy of Sciences, Beijing 100190, China}

\author{ G. Q. Zhang}
\affiliation{Shanghai Institute of Applied Physics, Chinese
Academy of Sciences, Shanghai 201800, China}
\author{ X. G. Cao}
\affiliation{Shanghai Institute of Applied Physics, Chinese
Academy of Sciences, Shanghai 201800, China}
\affiliation{Kavli Institute for Theoretical Physics, Chinese Academy of Sciences, Beijing 100190, China}
\author{ D. Q. Fang}
\affiliation{Shanghai Institute of Applied Physics, Chinese
Academy of Sciences, Shanghai 201800, China}
\affiliation{Kavli Institute for Theoretical Physics, Chinese Academy of Sciences, Beijing 100190, China}
\author{ H. W. Wang}
\affiliation{Shanghai Institute of Applied Physics, Chinese
Academy of Sciences, Shanghai 201800, China}
\author{ J. Xu}
\affiliation{Shanghai Institute of Applied Physics, Chinese
Academy of Sciences, Shanghai 201800, China}

\date{ \today}

\begin{abstract}

The isoscalar giant monopole resonance (ISGMR) in Sn isotopes and
other nuclei is  investigated in the framework of the
isospin-dependent quantum molecular dynamics (IQMD) model. The
spectrum of GMR is calculated by taking the root-mean-square (RMS)
radius of a nucleus as its monopole moment. The peak energy, the
full width at half maximum (FWHM), and the strength of GMR extracted
by a Gaussian fit to the spectrum have been studied. The GMR peak
energies for Sn isotopes from the calculations using a mass-number
dependent Gaussian wave-packet width $\sigma_r$ for nucleons are
found to be overestimated and show a weak dependence on the mass
number compared with the experimental data. However, it is found
that experimental data of the GMR peak energies for $^{56}$Ni,
$^{90}$Zr, and $^{208}$Pb as well as Sn isotopes can be nicely
reproduced after taking into account the isospin dependence in
isotope chains in addition to the mass number dependence of
$\sigma_r$ for nucleons in the IQMD model calculation.

\end{abstract}


\maketitle

\section{Introduction}

The isoscalar giant monopole resonance (ISGMR), known as the
so-called "breathing mode", is one of the collective modes of
nuclei. In the past decades, GMR was extensively studied both
theoretically~\cite{Civitarese-prc1991,Agrawal-prc2003,Colo-prc2004,Todd-Rutel-prl2005,Piekarewicz-prc2007,
Piekarewicz-prc2009,Tselyaev-prc2009,Khan-prc2009a,Khan-prc2009b,Khan-prl2012}
and
experimentally~\cite{Shlomo-prc1993,Li-prl2007,Grag-npa2007,Li-prc2010,Grag-appb2011}.
Especially, a strong correlation between the peak energy of GMR
and the nuclear incompressibility $K_0$ at the nuclear saturation
density was found~\cite{Blaizot-npa1995}. The studies from
both relativistic and non-relativistic models have reached a
consensus on the value of the nuclear incompressibility at
$K_0\sim240\pm10$
MeV~\cite{Agrawal-prc2003,Colo-prc2004,Todd-Rutel-prl2005}.

Recently, GMR along the Sn isotopic chain was studied
experimentally~\cite{Li-prl2007,Grag-npa2007,Li-prc2010}. From the
analysis based on the GMR data for Sn isotopes, the asymmetry term
of the nuclear incompressibility was constrained, i.e.,
$K_{\tau}\sim-550\pm100$ MeV~\cite{Li-prl2007}. Similar analysis on
the GMR data in the Cd isotopes gave a preliminary value of
$K_{\tau}\sim-555\pm 75$
MeV~\cite{Patel}.
 Comparison
between the experimental data and the theoretical results has
indicated that models which can reproduce the peak energies of GMR
in $^{90}$Zr, $^{144}$Sm, and $^{208}$Pb overestimate those in Sn
isotopes. This realization leaves a puzzling question: "why is Tin
so soft?"~\cite{Piekarewicz-prc2007}. In
Ref.~\cite{Civitarese-prc1991}, the effect of pairing correlations
on the peak energy of GMR was considered. However, the result
that the peak energies of GMR in Sn isotopes are shifted by about
100-150 KeV compared to the case without pairing correlations was
insufficient to explain the experimental data. In
Ref.~\cite{Piekarewicz-prc2009}, a hybrid model with a small nuclear
incompressibility of $K_0=230$ MeV as FSUGold and a stiff symmetry
energy as NL3 was built. Although the improvement in the
description for the experimental data of Sn isotopes was significant
and unquestionable, the hybrid model still underestimated the peak
energy of GMR in $^{208}$Pb by almost 1 MeV. The authors of
Ref.~\cite{Piekarewicz-prc2009} also suggested that the rapid
softening with neutron excess predicted by the hybrid model might be
unrealistic. More details of the discussion on this anomaly can be
found in
Refs.~\cite{Civitarese-prc1991,Piekarewicz-prc2009,Tselyaev-prc2009,Khan-prc2009a,Khan-prc2009b,Li-prc2010}.

In the previous works, our group have applied the isospin-dependent
quantum molecular dynamics (IQMD) model to study the dynamical
dipole emission in fusion reactions~\cite{Wu-prc2010} and giant
dipole resonances (GDR) as well as pygmy dipole resonances (PDR) in
Ni isotopes by Coulomb excitations~\cite{Tao-prc2013}. In the
present work, we will investigate GMR in Sn isotopes within a
similar framework. We will show that using a new function of the
Gaussian wave-packet width, which takes the isospin dependence into
account, we are able to reproduce very well the GMR peak energies
for $^{56}$Ni, $^{90}$Zr, and $^{208}$Pb as well as Sn isotopes.

The paper is organized as follows. Section~\ref{sec-2} gives a brief
introduction of the IQMD model as well as the formalism for GMR in
the IQMD framework. Results and discussions are presented in
Sec.~\ref{sec-3}, where effects from the impact parameter, the
incident energy, the equation of state (EOS), the symmetry energy,
and the width of the Gaussian wave packet used in the IQMD model on
GMR are investigated. A summary is given in Sec.~\ref{sec-4}.

\section{Model and Formalism}
\label{sec-2}
\subsection{Brief description of IQMD model}

The IQMD model, which is based on the QMD model, is a kind of
Monte-Carlo transport
model~\cite{Aichelin-pr1991,Hartnack-epja1998,Hartnack-npa1989,Ma-prc2006,Kumar-prc2010,Zhang-prc2011,Wang-nst2013}.
The wave function of each nucleon is represented by a Gaussian form:
\begin{equation}
\label{eq:gauspack}
  \phi_i(\vec{r},t) = \frac{1}{{(2\pi L)}^{3/4}}
\exp\left[-\frac{{(\vec{r}-
\vec{r_i}(t))}^2}{{(2\sigma_r)}^2}-\frac{i\vec{r} \cdot
\vec{p_i}(t)}{\hbar}\right].
\end{equation}
In the above, $\sigma_r$ is the width parameter for the Gaussian
wave-packet, and its value depends on the size of the reacting
system to keep some quantum effect of nucleons. We will see in the
following that the influence of the width on GMR should be treated
carefully to reproduce the experimental results. $\vec{r_i}(t)$ and
$\vec{p_i}(t)$ are the position and momentum coordinates of the
$i$th nucleon. After performing variation method, the equations of
motion, i.e., the time evolution of the mean position $\vec{r_i}(t)$
and momentum $\vec{p_i}(t)$, are found to be
\begin{equation}
\dot{\vec{p}}_i = - \frac{\partial \langle H \rangle}{\partial
\vec{r}_i} \quad {\rm and} \quad \dot{\vec{r}}_i = \frac{\partial
\langle H \rangle}{\partial \vec{p}_i}.
\end{equation}
$\langle H \rangle$ is the total Hamiltonian of the system
\begin{equation}
\label{eq:hamiltonian}
\langle H \rangle = \langle T \rangle + \langle V \rangle,
\end{equation}
where the $\langle T \rangle$ is the kinetic contribution, and
$\langle V \rangle$ is the potential contribution
\begin{equation}
\label{meanfield}
\langle V \rangle = \frac{1}{2} \sum_{i} \sum_{j \neq i}
 \int f_i(\vec{r},\vec{p},t) \,
V^{ij}  f_j(\vec{r}\,',\vec{p}\,',t)\, d\vec{r}\, d\vec{r}\,'
d\vec{p}\, d\vec{p}\,'.
\end{equation}
In the above, the Wigner distribution function $f_i
(\vec{r},\vec{p},t)$, which is the phase-space density of the $i$th
nucleon, is obtained by applying the Wigner transformation on the
single nucleon wave function
\begin{equation} \label{wignerfuction}
 f_i (\vec{r},\vec{p},t) = \frac{1}{\pi^3 \hbar^3 }
 \exp\left[-\frac{(\vec{r} - \vec{r}_{i} (t) )^2}{2{\sigma_r}^2}
-\frac{2{\sigma_r}^2(\vec{p} - \vec{p}_{i} (t) )^2}{\hbar^2}
\right],
\end{equation}
and $V^{ij}$ the two-body interaction including the contact
Skyrme-type interaction, the finite-range Yukawa potential, the
momentum-dependent interaction (MDI), the isospin-dependent
interaction, and the Coulomb interaction
\begin{eqnarray}
 \label{vijdef}
V^{ij} &=& V^{ij}_{\rm Skyrme} + V^{ij}_{\rm Yuk} + V^{ij}_{\rm asy}
+ V^{ij}_{\rm mdi} +
           V^{ij}_{\rm Coul} \nonumber \\
       &=& t_1 \delta (\vec{r} - \vec{r}^\prime) +
           t_2 \rho^{\sigma-1}(\vec{r}) \delta (\vec{r} - \vec{r}^\prime)\nonumber\\
       &+& t_3 \frac{\exp[-|\vec{r}-\vec{r}^\prime|/\mu]}{
               |\vec{r}-\vec{r}^\prime|/\mu} + t_6 \rho^{\gamma-1}(\vec{r})
 T_{3i} T_{3j} \delta(\vec{r} - \vec{r}^\prime) \nonumber \\
       &+& t_4\hbox{ln}^2 [1+t_5(\vec{p}-\vec{p}^\prime)^2]
               \delta (\vec{r} -\vec{r}^\prime) +
           \frac{Z_i Z_j e^2}{|\vec{r}-\vec{r}^\prime|},
\end{eqnarray}
where $Z$ is the charge of the nucleon, and $t_1...t_6$ and $\mu$
are the parameters to fit the empirical properties of nuclear matter
as well as nuclei.

In the following, we will give the expressions for the Skyrme
potential, the momentum-dependent potential, and the symmetry
potential to ease discussions for the GMR results. The Skyrme
potential is
\begin{equation}
\label{skyene} U_{\rm Sky} = \alpha u + \beta u^{\sigma},
\end{equation}
where $u=\rho_{\rm int}/\rho_{0}$ is the reduced density with
$\rho_0=0.16$ fm$^{-3}$ being the nuclear saturation density and
$\rho_{\rm int}=\sum{\rho_{ \rm int}^i(\vec{r})}$ with
\begin{equation} \label{rhoint}
\rho_{\rm int}^i(\vec{r}) = \frac{1}{(4\pi {\sigma_r}^2)^{3/2}}
\sum_{j \neq i} {\rm e}^{\displaystyle
-(\vec{r}-\vec{r_{j}})^2/(4{\sigma_r}^2) }
\end{equation}
being the interaction density of the $i$th nucleon. We will in the
following denote $\rho_{\rm int}$ as $\rho$ for simplicity.
$\alpha$, $\beta$, and $\sigma$ are the Skyrme parameters related to
the isoscalar EOS of bulk nuclear matter. The momentum dependent
potential, which is optional in the IQMD model, can be expressed as
\begin{equation}
\label{mdipar}
U_{\rm mdi} = \frac{\rho_{\rm int}}{\rho_0}\int d\vec{p}\,'g_j(\vec{p}\,',t)
\delta \cdot ln^2[\epsilon(\vec{p}-\vec{p}\,')^2+1],
\end{equation}
where $g_j(\vec{p},t)=\frac{1}{(\pi\hbar)^{3/2}}
 \exp\left[-\frac{2{\sigma_r}^2(\vec{p} - \vec{p}_{j} (t))^2}{\hbar^2}
\right]$ is the momentum density distribution function of nucleon, $\delta$ = 1.57 MeV and $\epsilon$ = 500 (GeV/c)$^{-2}$ are
taken from the measured energy dependence of the proton-nucleus
optical potential
\cite{Aichelin-pr1991,Hartnack-epja1998,Hartnack-npa1989,Ma-prc2006,Kumar-prc2010}.
The isospin asymmetry potential can be also calculated from
Eq.(\ref{meanfield})
\begin{eqnarray}
U_{\rm sym}&=&\frac{C_{\rm sym}}{2}[(\gamma-1)
u^{\gamma}\delta^2\pm2 u^{\gamma}\delta], \label{eq04}
\end{eqnarray}
where $\delta=(\rho_{\rm n}-\rho_{\rm p})/\rho$ is the local isospin
asymmetry from the contribution of all the other nucleons, and the
symbol "$+(-)$" is for neutrons (protons), $\gamma$ is the stiffness
parameter of the symmetry potential (energy), and $C_{\rm sym}$ is
the potential contribution of the symmetry energy at saturation
density.

The nuclear incompressibility is calculated from the second-order
derivative of the binding energy per nucleon
\begin{equation}
K_0 = 9 \rho_0^2 \frac{\partial^2}{\partial \rho^2} \left(
\frac{E}{A}\right)_{\rho=\rho_0}.
\end{equation}
Table~\ref{table1} gives different parameter sets for the Skyrme
potential with and without the momentum-dependent potential, leading
to the nuclear incompressibility of $200$ MeV and $380$ MeV. As
mentioned in the introduction, although the latest experimental
analysis leads to $K_0$ $\approx$ 240 MeV, we will use these extreme
values to illustrate the sensitivity of the GMR peak energy on the
nuclear incompressibility as well as the momentum dependence of the
nuclear potential based on the IQMD model.

\begin{table}[htbp]
\caption{The parameters $\alpha$, $\beta$, and $\sigma$ for
different EOSs.} \label{table1} \centering
\begin{tabular}{p{60pt}p{42pt}p{42pt}p{42pt}p{42pt}}
\hline \hline
 & $K_{0}$ & $\alpha$ & $\beta$ & $\sigma$ \\
 & (MeV) & (MeV) & (MeV) &  \\
\hline
soft & 200 & -356 & 303 & 7/6 \\
soft+MDI & 200 & -390.1 & 320.3 & 1.14 \\
hard & 380 & -124 & 70.5 & 2 \\
hard+MDI & 380 & -129.2 & 59.4 & 2.09 \\
\hline \hline
\end{tabular}
\end{table}

\subsection{Calculation method of Giant Monopole Resonance}

\begin{figure}[htbp]
\resizebox{8.6cm}{!}{\includegraphics{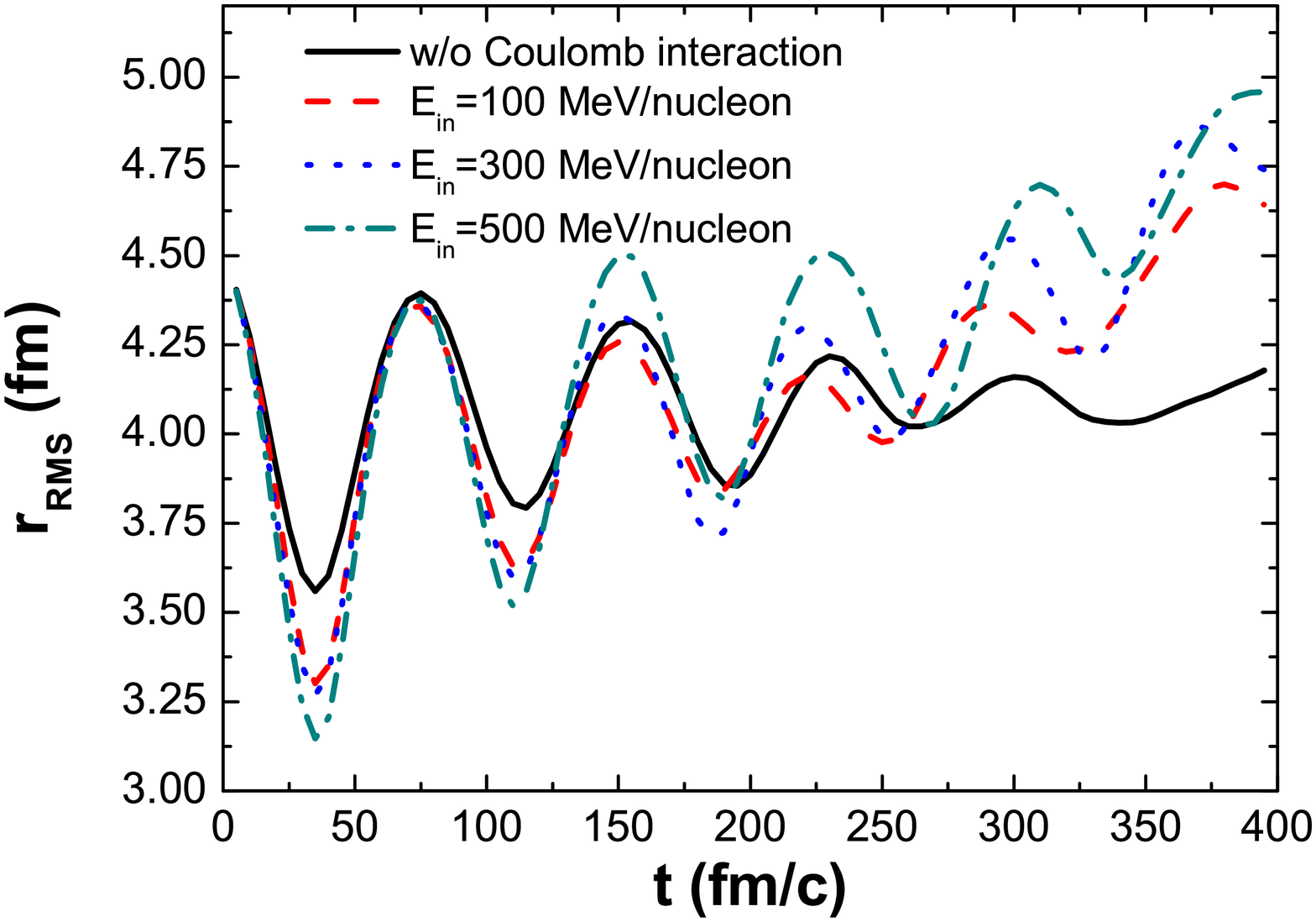}} \vspace{-0.4cm}
\caption{(Color online) Time evolution of the calculated RMS radius
of $^{112}$Sn without Coulomb interaction and reacting on the
$^{208}$Pb target at different beam energies of 100, 300, and 500
MeV/nucleon, respectively, with impact parameter of 30 fm.} \label{rms}
\end{figure}

In the present framework, we first pick up a stable initial density
distribution for a concerned nucleus, e.g., $^{112}$Sn, as is done
for most QMD model studies. This density distribution is generally
not the ground state for the nuclear interaction used, so the
nucleus suffers from collective oscillation in its excited
state~\cite{Gaitanos-prc2010}, among them is the GMR mode. As GMR is
a compression mode in radial direction, we take the root-mean-square
(RMS) radius of the nucleus as its monopole moment
$DR_{\textrm{GMR}}(t)$ at each time
step~\cite{Baran-npa2001,Gaitanos-prc2010}, as is shown by the solid
line in Fig.~\ref{rms}. One sees that the RMS radius shows good
oscillation structure, i.e., the GMR mode. However, the oscillation
damps quickly due to the dissipation effect from both the mean-field
potential and the nucleon-nucleon
scatterings~\cite{Gaitanos-prc2010}. It is noteworthy that the
period and the decay of the GMR oscillation leads respectively to
the peak energy and the width of the GMR spectrum, which can still
be hardly reproduced simultaneously~\cite{Gaitanos-prc2010}. In the
present study, we use $^{208}$Pb as a target and the concerned
nucleus as the projectile. In this way, the strength, period, and
damping of the GMR oscillation are thus modified by the Coulomb
interaction between the projectile and the target, as shown in
Fig.~\ref{rms}. It is seen that with Coulomb interaction, the RMS
radii have a larger amplitude and a little longer oscillation period
with the increasing of the beam energy and become larger on average
in the later stage. Overall, a good oscillation behavior of the RMS
radius in relative long time scale is the key to form a GMR mode. Of
course, considering the stability of time evolution of the RMS
radius in the present model, we calculate the GMR spectra by 200
fm/c.

By applying the Fourier transformation to the second-order
derivative of $DR_{\textrm{GMR}}(t)$ with respect to time
\begin{equation}
DR''(\omega)=\int_{t_0}^{t_{\textrm{max}}}DR''_{\rm
GMR}(t)e^{i\omega t}dt,
\end{equation}
 one can get the spectrum of probability for energy
$E_{\gamma}=\hbar\omega$ as follows
\begin{equation}
\frac{dP}{dE_{\gamma}}=\frac{2e^2}{3\pi \hbar c^3E_{\gamma}}\lvert DR''(\omega)\rvert ^2.
\end{equation}
As mentioned above, we set the stopping time of the monopole moment
($t_{max}$) as 200 fm/c in the present calculation. From a Gaussian
fit to the spectrum of GMR, one can get the peak energy
$E_{\gamma}^c$, the FWHM $\Gamma_{\gamma}^c$, and the strength
$S_{\gamma}^c$ of GMR.

\section{Results and Discussions}
\label{sec-3}
\subsection{GMR spectrum comparison}

\begin{figure}[htbp]
\resizebox{8.6cm}{!}{\includegraphics{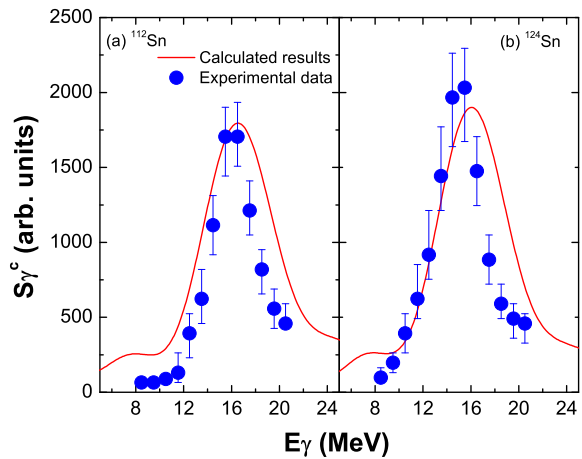}} \vspace{-0.4cm}
\caption{(Color online) The calculated results of GMR in $^{112}$Sn
and $^{124}$Sn compared with the experimental data. Note that the
condition to induce the GMR is different for the experimental data
and the present calculation (see texts for more details). The blue
circles with error bar are the experimental data from inelastic
scattering \cite{Li-prl2007}, and the red line is the result from
the IQMD calculation.} \label{sn-exp-cal}
\end{figure}

The GMR spectra for $^{112}$Sn and $^{124}$Sn from our calculations
are compared with the experimental data from Ref.~\cite{Li-prl2007}
in Fig.~\ref{sn-exp-cal}. Note that there are one major different
point which we should mention in this comparison between our
calculation and the data. In Ref.~\cite{Li-prl2007}, the GMR data
are taken from the excited Sn nucleus by inelastic scattering of
400-MeV $\alpha$ particles at extremely forward angles. However, in
our calculation, the GMR comes from the excited oscillation of Sn
nucleus as the initial density distribution is not the ground state
of the nuclear interaction used. The reason is based on the
following consideration: (1) it is not an easy task to treat
inelastic scattering of $\alpha$-particles in our IQMD model even
though the data is available; (2) the peak energy and the FWHM are
determined by the intrinsic properties of the nucleus and
independent of how the GMR mode is excited. Consequently, we have to
take a compromise for the comparison, i.e., taking the GMR for the
same excited nucleus but with different reaction mechanism, with the
strength $S_{\gamma}^c$ of GMR from our calculation scaled by that
from the experimental data. In this background, we introduce the
parameters used in our calculations as follows: incident energy
$E_{\rm in} = 386$ MeV/nucleon, impact parameter $b = 30$ fm, the
soft EOS with MDI, $C_{\rm sym} = 35.2$ MeV, and $\gamma=1$.
Although the condition of our calculation is different from that of
the experiment in Ref.~\cite{Li-prl2007} as mentioned above, one can
see that our results from the oscillation of excited nuclei modified
by the Coulomb interaction show a reasonable agreement with the
inelastic $\alpha$ scattering experimental data, i.e., giving a
similar peak energy but a slightly larger FWHM. The calculated
result of $^{112}$Sn shows a better agreement with the experimental
data than that of $^{124}$Sn. The results indicate that the IQMD
model is suitable for the study of GMR by considering the RMS radius
as its monopole moment.

\subsection{Systematic GMR analysis}

\begin{figure}[htbp]
\resizebox{8.6cm}{!}{\includegraphics{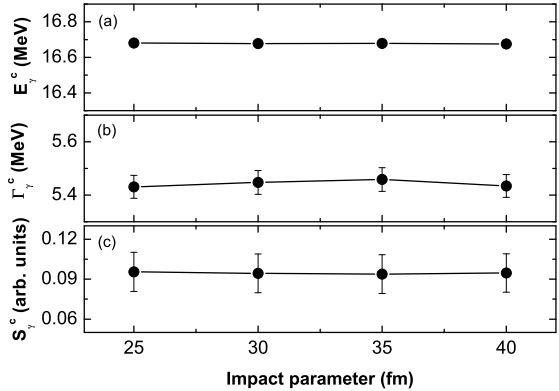}} \vspace{-0.4cm}
\caption{Impact parameter dependence of the peak energy
$E_{\gamma}^c$ (upper panel), the FWHM $\Gamma_{\gamma}^c$ (middle
panel), and the strength $S_{\gamma}^c$ (bottom panel) of GMR for
$^{112}$Sn. In the calculation, we use $E_{\textrm{in}}$ = 386
MeV/nucleon, $C_{\textrm{sym}}$ = 35.2 MeV, $\gamma$ = 1, and the
soft EOS with MDI.} \label{b-dep}
\end{figure}

\begin{figure}[htbp]
\resizebox{8.6cm}{!}{\includegraphics{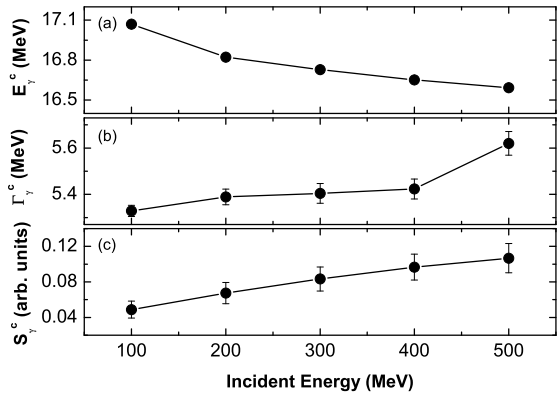}}
 \vspace{-0.4cm}
\caption{Same as Fig~\ref{b-dep} but for the incident energy
dependence of the GMR results for $^{112}$Sn.  In the calculation,
we use $b$ = 30 fm, $C_{\textrm{sym}}$ = 35.2 MeV, $\gamma$ = 1,
and the soft EOS with MDI. \label{E-dep}}
\end{figure}

The sensitivities of the peak energy, the FWHM, and the strength
obtained by a Gaussian fit to the GMR spectrum have been explored.
The sensitivity of these quantities to the impact parameter for
$^{112}$Sn is given in Fig.~\ref{b-dep}. It shows that the GMR
results do not change much with the increasing impact parameter.
This is understandable as the effect of Coulomb interaction is not
affected by much when the impact parameter changes from 25 fm to 40
fm.
Figure~\ref{E-dep} shows the incident energy dependence of GMR
results for $^{112}$Sn. With the increase of the incident energy,
the peak energy of GMR decreases, while the FWHM and the strength of
GMR increase. This behavior can be understood by the oscillation of
RMS radii at different beam energies as shown in Fig.~\ref{rms}. As
we know, the longer oscillation period corresponds to the lower
frequency, i.e. lower energy, while the higher amplitude corresponds
to the larger strength. Fig.~\ref{rms} tells us that with the
increasing of beam energy, GMR monopole moment has a little longer
period but a larger amplitude, which results in a decreasing peak
energy and an increasing strength of GMR as shown in
Fig.~\ref{E-dep}. From the above discussions, it is seen that
although the GMR oscillation is already there for a nucleus alone,
it can be slightly modified by the Coulomb interaction with
different incident energies and impact parameters.

\begin{figure}[htbp]
\resizebox{8.6cm}{!}{\includegraphics{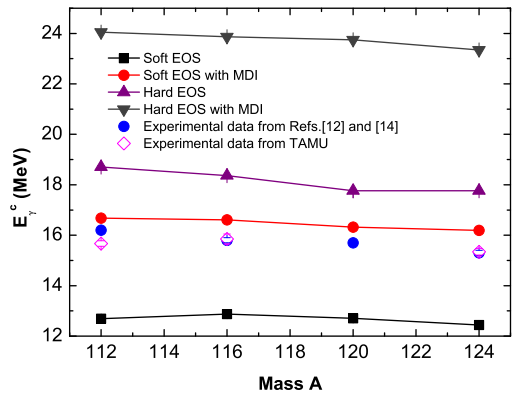}} \vspace{-0.4cm}
\caption{(Color online) The peak energies of GMR in Sn isotopes when
different EOS parameters are used. In the calculation, we use
$E_{\textrm{in}}$ = 386 MeV/nucleon, $b$ = 30 fm, $C_{\textrm{sym}}$
= 35.2 MeV, and $\gamma$ = 1. The blue circles with error bar are
the experimental data from Ref.~\cite{Li-prl2007}, and the open
diamonds with error bar are the experimental data from
Refs.~\cite{Youngblood-prc2004,Lui-prc2004}.} \label{eos-dep}
\end{figure}

Many previous works indicated that the EOS associated with the
nuclear incompressibility $K_{0}$ has an important influence on GMR,
i.e., the peak energy of GMR can be used to constrain $K_{0}$. By
adjusting the parameters of EOS in Table.~\ref{table1} used in the
IQMD model, the EOS dependence of GMR parameters for $^{112}$Sn can
be explored. The sensitivity of the GMR results to the EOS are
illustrated in Fig.~\ref{eos-dep}, where a significant dependence of
the GMR peak energy on the EOS can be seen. In general, the hard EOS
gives a higher peak energy than the soft one, and so does the EOS
with MDI. In this sense, a soft EOS with MDI can reproduce the
results from a hard EOS without MDI. Similar results on giant or
pygmy dipole resonance are seen within the same model
\cite{Tao-prc2013}. By comparing the calculated results with the
experimental data~\cite{Li-prl2007,Youngblood-prc2004,Lui-prc2004},
one can see that the soft EOS with MDI shows the best agreement with
the experimental data.

\begin{figure}[htbp]
\resizebox{8.6cm}{!}{\includegraphics{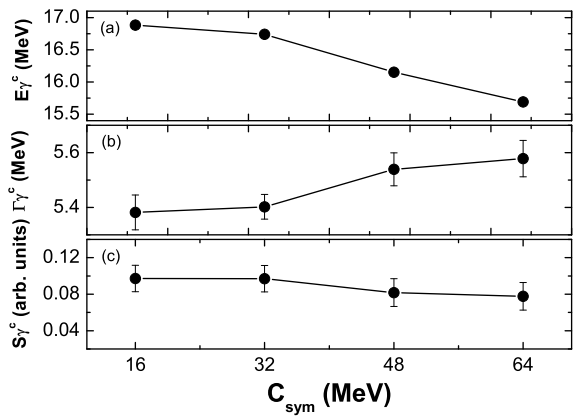}} \vspace{-0.4cm}
\caption{Same as Fig~\ref{b-dep} but for C$_{\textrm{sym}}$
dependence of GMR results for $^{112}$Sn. In the calculation, we use
$E_{\textrm{in}}$ = 386 MeV/nucleon, $b$ = 30 fm, $\gamma$ = 1, and
the soft EOS with MDI.} \label{csym-dep}
\end{figure}

\begin{figure}[htbp]
\resizebox{8.6cm}{!}{\includegraphics{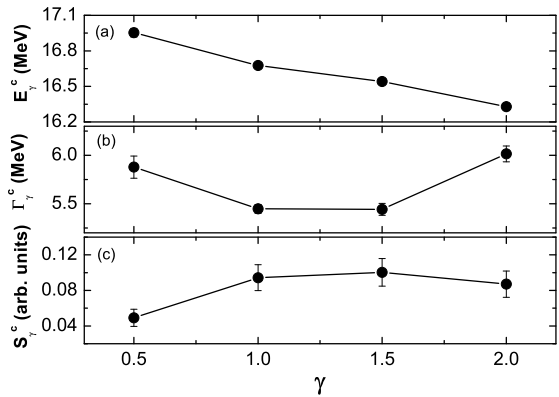}} \vspace{-0.4cm}
\caption{Same as Fig~\ref{b-dep} but for the $\gamma$ dependence of
GMR results for $^{112}$Sn. In the calculation, we use
$E_{\textrm{in}}$ = 386 MeV/nucleon, $b$ = 30 fm, $C_{\textrm{sym}}$
= 35.2 MeV, and the soft EOS with MDI. \label{gamma-dep}}
\end{figure}

The GMR results may also be affected by the nuclear symmetry energy
which is important in understanding the structure of neutron- or
proton-rich nuclei and the reaction dynamics of heavy-ion collisions
\cite{Li,Jiang}. Again we use the extreme values of $C_{\rm sym}$
and $\gamma$ to illustrate the sensitivity of the GMR results to the
nuclear symmetry energy. The dependence of the GMR results on the
parameter $C_{\rm sym}$ is shown in Fig.~\ref{csym-dep}. When
$C_{\rm sym}$ changes from 16 MeV to 64 MeV, the peak energy of GMR
shows a decreasing behavior, the FWHM of GMR increases, and the
strength of GMR slightly decreases.
The $\gamma$ dependence of the GMR results is shown in
Fig.~\ref{gamma-dep}. When $\gamma$ changes from 0.5 to 2, the peak
energy of GMR also decreases, while the FWHM and the strength of GMR
show a non-monotonical behavior.
To understand the dependence of the GMR peak energy on the symmetry
energy, we express it in the form of~\cite{Stringari-plb1982}:
\begin{equation}
E_{\gamma}^c=\hbar\sqrt{\frac{K_A}{{\rm m}\langle r^2\rangle}},
\end{equation}
where $m$ is the nucleon mass, $\langle r^2\rangle$ is the
ground-state mean square radius, and $K_A$, which is the
incompressibility of a nucleus with mass A, can be written
as~\cite{Li-prl2007}:
\begin{eqnarray}
K_A&\sim &K_V(1+c_VA^{-1/3})+K_{\tau}[(N-Z)/A]^2 \nonumber \\
&&+K_CZ^2A^{-4/3}, \label{eq-ka}
\end{eqnarray}
with $N$ and $Z$ the neutron and proton number, and $K_V$ as well as
$c_V$, $K_\tau$, and $K_C$ the coefficients for the volume,
asymmetry, and Coulomb contributions, respectively. In the analysis
of the symmetry energy effects on the GMR results where all the
other parameters have been fixed, $K_A$ increases with the
increasing asymmetry incompressibility $K_{\tau}$, with the latter
expressed as~\cite{Chen-prc2009}
\begin{eqnarray}
K_{\tau}=K_{\rm sym}-6L_s-\frac{J_0}{K_0}L_s. \label{Ktau}
\end{eqnarray}
The slope parameter $L_s$ and the curve parameter $K_{\rm sym}$ at
saturation density can be calculated from the expression of the
symmetry energy as
\begin{eqnarray}
L_s&=&25+\frac{3}{2}C_{\rm sym}\cdot\gamma~{\rm (MeV)}, \nonumber \\
K_{\rm sym}&=&-25+\frac{9}{2}C_{\rm sym}\gamma(\gamma-1) ~{\rm
(MeV)}. \label{eq-slope-curve}
\end{eqnarray}
$K_0$ and $J_0$ are related to the isoscalar part of the equation of
state $E_0(\rho)$ in the form of
\begin{eqnarray}
K_0 &=& 9\rho_0^2 \left(\frac{d^2E_0}{d\rho^2}\right)_{\rho=\rho_0},\\
J_0 &=& 27\rho_0^3
\left(\frac{d^3E_0}{d\rho^3}\right)_{\rho=\rho_0}.
\end{eqnarray}
The dependence of $E_\gamma^c$ on $C_{\rm sym}$ and $\gamma$ in
Figs.~\ref{csym-dep} and \ref{gamma-dep} can be understood from the
above formulae.

\begin{figure}[htbp]
\resizebox{8.6cm}{!}{\includegraphics{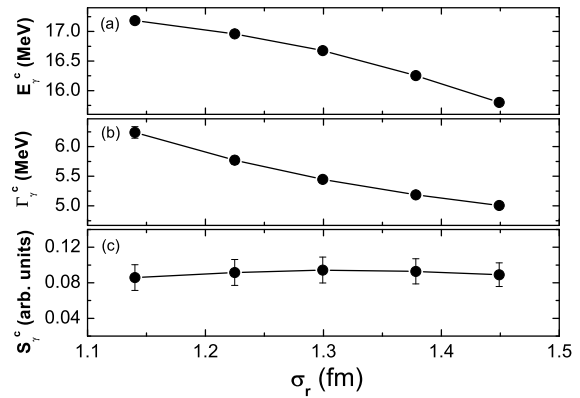}} \vspace{-0.4cm}
\caption{Same as Fig~\ref{b-dep} but for $\sigma_r$ dependence of
GMR results for $^{112}$Sn. In the calculation, we use
$E_{\textrm{in}}$ = 386 MeV/nucleon, $b$ = 30 fm, $C_{\textrm{sym}}$
= 35.2 MeV, $\gamma$ = 1, and the soft EOS with MDI.
\label{del-dep}}
\end{figure}

The Gaussian wave-packet width $\sigma_r$ (in Eq. (\ref{eq:gauspack})) for nucleons is a
parameter in the IQMD model indicating the interaction range between
nucleons. Previous studies have shown that the value $\sigma_r$ has
a large effect on GMR results~\cite{Furuta-prc2010}. In many
previous QMD model calculations, $\sigma_r$ is set to be a constant.
However, there were also some discussions related to the influence
of $\sigma_r$ on the dynamical results, e.g., flow,
multifragmentation, and pion and kaon production, etc.
\cite{Hartnack-epja1998,Gautam-jpg2010,Maruyama-prc1996,Feldmeier-npa1990,Feldmeier-npa1995,Wang-prc2002}.
Since in a finite system nucleons are localized within a potential
well, it is reasonable to make $\sigma_r$ related to the size of a
nucleus. For example, the width of the Gaussian wave packet is taken
to be $\sigma_r=1.04$ fm for Ca + Ca system and $\sigma_r=1.47$ fm
for Au+Au system in Ref.~\cite{Hartnack-epja1998}, while a
system-size-dependent $\sigma_r$ was presented in
Ref.~\cite{Wang-prc2002}. Actually, in some models like an Extended
Quantum Molecular Dynamics (EQMD) model~\cite{Maruyama-prc1996} and
a Fermionic Molecular Dynamics (FMD)
model~\cite{Feldmeier-npa1990,Feldmeier-npa1995}, $\sigma_r$ is
treated as a dynamical variable. In our IQMD model, we use a similar
mass-number dependence of $\sigma_r$ as in Ref.~\cite{Wang-prc2002}
\begin{equation}
\sigma_r=0.17A^{1/3}+0.48~{\rm (fm)}, \label{eq-sigma}
\end{equation}
where $A$ is the mass number of the system (projectile, target, or
compound system). In heavy-ion reactions, $\sigma_r$ for the
compound system is set to be the mean value of that for the
projectile and the target given by Eq.~(\ref{eq-sigma}). In the
present study, we just use $\sigma_r$ for the concerned nucleus,
i.e, the projectile. This is similar to the treatment in
Ref.~\cite{Wang-prc2002}, where the projectile and target have their
own $\sigma_r$ before they contact in the heavy-ion reaction
process, and after the contact, the projectile and the target
gradually melt into one system, and consequently all the particles
have a universal $\sigma_r$. Figure~\ref{del-dep} displays the
$\sigma_r$ dependence of the GMR results. As $\sigma_r$ increases,
the peak energy and the FWHM of GMR show a clear decreasing trend,
while the strength of GMR is not largely affected.

\begin{figure}[htbp]
\resizebox{8.6cm}{!}{\includegraphics{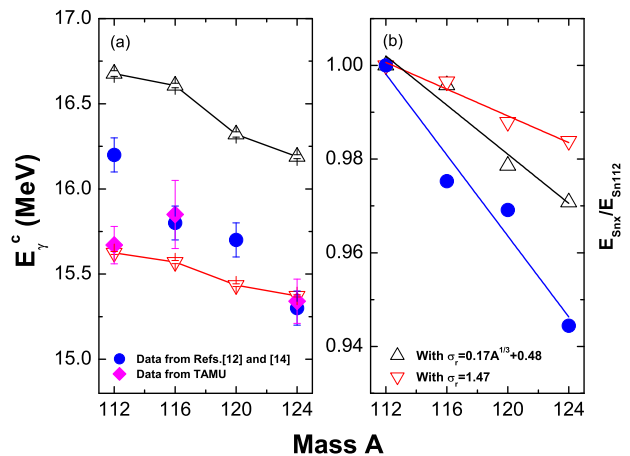}} \vspace{-0.4cm}
\caption{(Color online) Mass number dependence of the GMR peak
energies for Sn isotopes with a fixed (down-triangles) or a
mass-number dependent (up-triangle) $\sigma_r$. In the calculation,
we use $E_{\textrm{in}}$ = 386 MeV/nucleon, $b$ = 30 fm,
$C_{\textrm{sym}}$ = 35.2 MeV, $\gamma$ = 1, and the soft EOS with
MDI. In the left panel, the blue circles with error bar are the
experimental data from Ref.~\cite{Li-prl2007}, and the pink diamonds
with error bar are the experimental data from
Refs.~\cite{Youngblood-prc2004,Lui-prc2004}. In the right panel, the
peak energies of GMR in Sn isotopes are divided by those in
$^{112}$Sn for better illustrating the mass-number dependence, and
the straight lines are plotted to guide eyes. \label{del-com}}
\end{figure}

The influence of different forms of $\sigma_r$ on the peak energy of
GMR has also been studied. Figure~\ref{del-com} shows the calculated
peak energies of GMR in Sn isotopes with the fixed $\sigma_r=1.47$
fm or the variational $\sigma_r$ given by Eq.~(\ref{eq-sigma})
together with the experimental
data~\cite{Li-prl2007,Youngblood-prc2004,Lui-prc2004}, respectively.
One can see that the mass-number dependent $\sigma_r$ gives a larger
peak energy and shows a stronger mass number dependence of the GMR
peak energy than the constant $\sigma_r$. However, it still gives a
weaker mass-number dependence of the GMR peak energy in comparison
with the experimental data as seen from the right panel of
Fig.~\ref{del-com}. Results from other theoretical
studies~\cite{Colo-prc2004,Piekarewicz-prc2007} are also plotted for
comparison. One can see that they overestimate the peak energies of
GMR by about $0.3-1$ MeV in Sn isotopes, and the mass-number
dependence of the GMR peak energy seems also weaker than the
experimental data.

\subsection{Fitting GMR peak energies with $\sigma_r$}

\begin{figure}[htbp]
\resizebox{8.6cm}{!}{\includegraphics{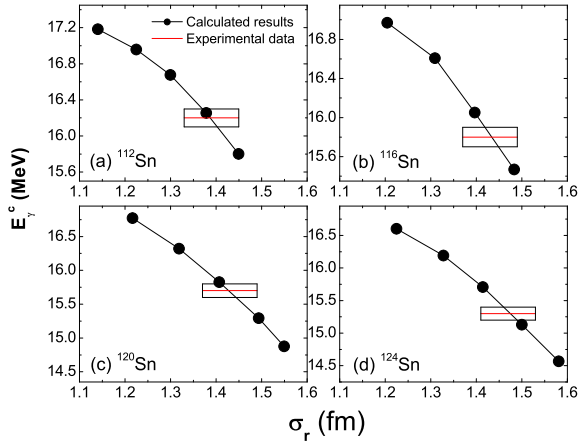}} \vspace{-0.4cm}
\caption{(Color online) Fit of $\sigma_r$ by the peak energies of
GMR in $^{112}$Sn, $^{116}$Sn, $^{120}$Sn, and $^{124}$Sn. In the
calculation, we use $E_{\textrm{in}}$ = 386 MeV/nucleon, $b$ = 30
fm, $C_{\textrm{sym}}$ = 35.2MeV, $\gamma$ = 1, and the soft EOS
with MDI. The red line with the black rectangle representing the
error is the experimental data from Ref.~\cite{Li-prl2007}.
\label{snx-dep}}
\end{figure}

From the above discussions, it is seen that introducing only the
mass-number dependence to the width of the Gaussian wave packet for
nucleons in the IQMD model is not sufficient to reproduce the
experimental results of the GMR peak energies for Sn isotopes.
However, considering the correlation between $\sigma_r$ and the peak
energy of GMR, we can fit the values of $\sigma_r$ with the
experimental data for different Sn isotopes. Figure~\ref{snx-dep}
shows such a fit of $\sigma_r$ for $^{112}$Sn, $^{116}$Sn,
$^{120}$Sn, and $^{124}$Sn using the experimental results of GMR
peak energies from Ref.~\cite{Li-prl2007}. It is seen that with the
increasing mass and isospin asymmetry along the Sn isotope line, the
GMR peak energy decreases while $\sigma_r$ from fitting the GMR peak
energy increases.

\begin{figure}[htbp]
\resizebox{8.6cm}{!}{\includegraphics{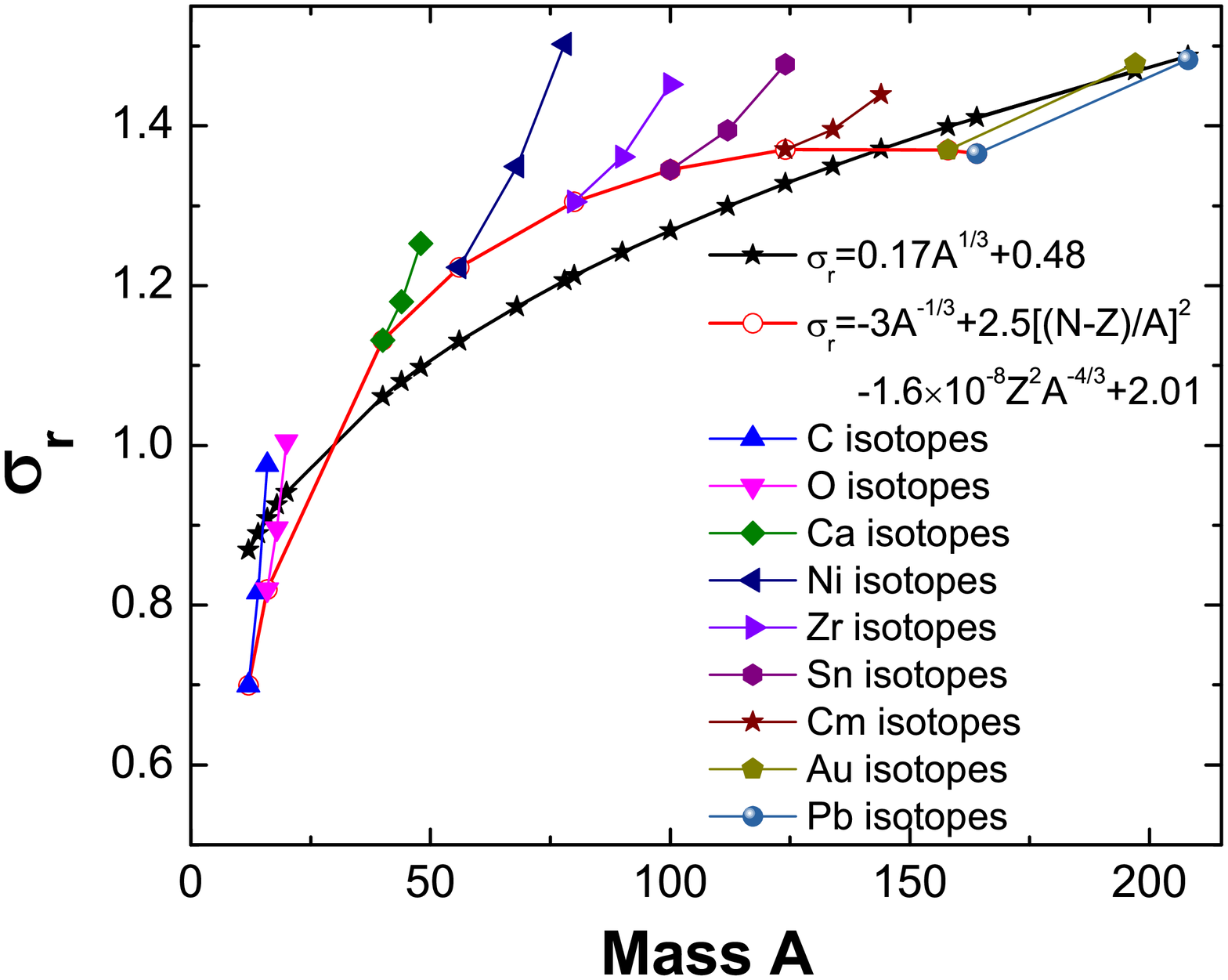}} \vspace{-0.4cm}
\caption{(Color online) Mass number dependence of the new function
of $\sigma_r$ given by Eq.~(\ref{eq-del-iso}) compared with that
given by Eq.~(\ref{eq-sigma}). The black line is the mass-number
dependence given by Eq.~(\ref{eq-sigma}). The red line is the one
given by Eq.~(\ref{eq-del-iso}) for symmetric nuclei, and other
lines are given by Eq.~(\ref{eq-del-iso}) for each isotopic chain.
\label{del-iso}}
\end{figure}

The above study gives us some hints that to reproduce reasonably
well the experimental data of the GMR peak energies for Sn isotopes,
$\sigma_r$ may depend not only on the mass number but also on the
isospin asymmetry of the nucleus. On the other hand, since the GMR
peak energy is strongly correlated with the nucleus
incompressibility $K_A$~\cite{Li-prl2007}, it is reasonable to
assume that $\sigma_r$ has a functional form similar to that of
$K_A$, i.e.,
\begin{equation}
\sigma_r=aA^{-1/3}+b[(N-Z)/A]^2+cZ^2A^{-4/3}+d.
\end{equation}
By fitting the experimental data of the GMR peak energies for Sn
isotopes~\cite{Li-prl2007} with the above functional form, we obtain
the following expression for $\sigma_r$
\begin{eqnarray}
\sigma_r&=&-3A^{-1/3}+2.5[(N-Z)/A]^2 \nonumber \\
&&+1.6\times10^{-8}Z^2A^{-4/3}+2.01~{\rm (fm)}. \label{eq-del-iso}
\end{eqnarray}
The small coefficient for the charge number dependence indicates
that the optimized choice of $\sigma_r$ might be a quadratic
function of the nucleus isospin asymmetry. The new function of
$\sigma_r$ is compared with the old one (Eq.~(\ref{eq-sigma})) used
in the QMD model in Fig.~\ref{del-iso}, and it is seen that they are
quite different especially for isotope chains.

\begin{figure}[htbp]
\resizebox{8.6cm}{!}{\includegraphics{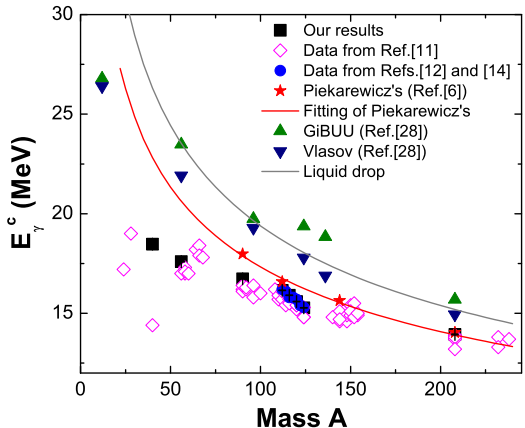}} \vspace{-0.4cm}
\caption{(Color online) Mass number dependence of the peak energy of
GMR with $\sigma_r$ given by Eq.~(\ref{eq-del-iso}). The blue
circles with error bar are the experimental data from
Ref.~\cite{Li-prl2007}, and the open diamonds with error bar are the
experimental data from Ref.~\cite{Shlomo-prc1993} and references
therein. The red line is a fitting of Piekarewicz's results with
$E_\gamma^c=69A^{-0.3}$ (MeV), and the gray line is a liquid drop
fitting with $E_\gamma^c=90A^{-1/3}$ (MeV). In the calculation, we
use $E_{\textrm{in}}$ = 386 MeV/nucleon, $b$ = 30 fm,
$C_{\textrm{sym}}$ = 35.2 MeV, $\gamma$ = 1, and the soft EOS with
MDI. } \label{a-dep-iso}
\end{figure}

Equation~(\ref{eq-del-iso}) has also been extended in the
calculation of GMR peak energies for $^{40}$Ca, $^{56}$Ni, and
$^{90}$Zr as well as $^{208}$Pb and the overall results are compared
with those from other theoretical
models~\cite{Piekarewicz-prc2009,Gaitanos-prc2010} as well as the
experimental data~\cite{Shlomo-prc1993,Li-prl2007} in
Fig.~\ref{a-dep-iso}. It is found that after introducing the isospin
dependence to the Gaussian wave-packet width fitted by the GMR peak
energies for Sn isotopes of intermediate nucleus mass, our results
agree with the experimental data for light or heavy nuclei such as
$^{56}$Ni, and $^{90}$Zr as well as $^{208}$Pb much better than
others in the literature, which overestimate the GMR peak energies.
For the even lighter nucleus such as $^{40}$Ca, although our result
follows the same trend of mass-number dependence for intermediate
and heavy nuclei, it overestimates the GMR peak energy compared with
the experimental data. Different from the cases of intermediate and
heavy nuclei, the GMR peak energies from the experimental studies
fluctuate with mass number for light nuclei. This is due to the
increasing effect of the shell structure as well as the paring
correlation for smaller nuclei, which has already been beyond the
limit of the framework of the IQMD model and the other transport
models with only mean-field potentials.

\section{Summary}
\label{sec-4}

In this work, we have applied the IQMD model to investigate the
isoscalar giant monopole resonance (ISGMR) in Sn isotopes and other
nuclei. The collective oscillation, including the GMR mode, appears
for an initial density distribution for the concerned nucleus as it
is generally not the ground state for the nuclear interaction used.
This oscillation is further modified by the Coulomb interaction when
we take the concerned nucleus as the projectile and the $^{208}$Pb
nucleus as the target. We took the RMS radius of the concerned
nucleus as the monopole moment of GMR and calculated the spectrum of
GMR. Using a Gaussian fit to the spectrum, we calculated the peak
energy, the FWHM, and the strength of GMR. The sensitivity of these
GMR results to the parameters used in the IQMD model was discussed.
The GMR peak energy is found to slightly decrease with increasing
incident energy, while it is almost independent of the impact
parameter. It seems difficult to extract the information of the
symmetry energy from the present study of GMR, as we found that GMR
is also sensitive to other parameters such as the isoscalar part of
the EOS and the Gaussian wave-packet width $\sigma_r$. As observed
previously, the EOS associated with the nuclear incompressibility
has an important influence on GMR. Comparing our results with the
experimental data, it is found that the soft EOS with MDI can give a
better fit to the experimental data. The studies of the systematic
evolution for Sn isotopes have shown that a widely used mass-number
dependent $\sigma_r$ overestimates the peak energies of GMR by about
1 MeV for Sn isotopes and leads to a weaker mass-number dependence
of the GMR peak energies compared with the experimental data. By
fitting the experimental data of the GMR peak energies for Sn
isotopes using the functional form of the nucleus incompressibility,
we obtain a new function of $\sigma_r$ with isospin dependence in
addition to mass-number dependence. Applying this new form of
$\sigma_r$ to the calculation of the GMR peak energies leads to a
good agreement with the experimental data for $^{56}$Ni, $^{90}$Zr,
and $^{208}$Pb, although it still overestimates the GMR peak energy
of $^{40}$Ca due to the lacking of effects from shell structure as
well as pairing correlation in the IQMD framework. It will be
interesting to check whether the width of the Gaussian wave packet
for nucleons obtained from the present study can give better
explanations for other observables from IQMD model calculations in
the future studies.

\section*{Acknowledgements}

This work was supported in part by the National Natural  Science
Foundation of China under Contract Nos. 11035009, 11220101005,
10979074, 11175231,11205230 the Major State Basic Research Development
Program in China under Contract No. 2013CB834405, the Knowledge
Innovation Project of Chinese Academy of Sciences under Grant No.
KJCX2-EW-N01, and the "100-talent plan" of Shanghai Institute of
Applied Physics under grant Y290061011 from the Chinese Academy of
Sciences.


\end{document}